\documentclass[aps,prb,twocolumn,superscriptaddress,showpacs]{revtex4}
\makeatletter

\newcommand{\Rmnum}[1]{\expandafter\@slowromancap\romannumeral #1@}
\makeatother

\usepackage{eurosym}
\usepackage{amsfonts}
\usepackage{amssymb}
\usepackage{amsmath}
\usepackage{graphicx}
\usepackage{color}
\begin{document}

\title{Enhanced thermoelectric performance of phosphorene by strain-induced band convergence}

\author{H. Y. Lv}

\affiliation{Key Laboratory of Materials Physics, Institute of Solid State Physics, Chinese Academy of Sciences, Hefei 230031, People's Republic of China}

\author{W. J. Lu}
\email[Corresponding author: ]{wjlu@issp.ac.cn}
\affiliation{Key Laboratory of Materials Physics, Institute of Solid State Physics, Chinese Academy of Sciences, Hefei 230031, People's Republic of China}

\author{D. F. Shao}

\affiliation{Key Laboratory of Materials Physics, Institute of Solid State Physics, Chinese Academy of Sciences, Hefei 230031, People's Republic of China}

\author{Y. P. Sun}
\email[Corresponding author: ]{ypsun@issp.ac.cn}
\affiliation{Key Laboratory of Materials Physics, Institute of Solid State Physics, Chinese Academy of Sciences, Hefei 230031, People's Republic of China}
\affiliation{High Magnetic Field Laboratory, Chinese Academy of Sciences, Hefei 230031, People's Republic of China}
\affiliation{University of Science and Technology of China, Hefei 230026, People's Republic of China}

\makeatletter


\begin{abstract}
The newly emerging monolayer phosphorene was recently predicted to be a promising thermoelectric material. In this work, we propose to further enhance the thermoelectric performance of phosphorene by the strain-induced band convergence. The effect of the uniaxial strain on the thermoelectric properties of phosphorene was investigated by using the first-principles calculations combined with the semi-classical Boltzmann theory. When the zigzag-direction strain is applied, the Seebeck coefficient and electrical conductivity in zigzag direction can be greatly enhanced simultaneously at the critical strain of 5\% where the band convergence is achieved. The largest $ZT$ value of 1.65 at 300 K is then achieved conservatively estimated by using the bulk lattice thermal conductivity. When the armchair-direction strain of 8\% is applied, the room-temperature $ZT$ value can reach 2.12 in the armchair direction of phosphorene. Our results indicate that strain-induced band convergence could be an effective method to enhance the thermoelectric performance of phosphorene.

\end{abstract}
\maketitle

\section{INTRODUCTION}

Thermoelectric materials, which can directly convert heat into electricity and vice versa, have attracted much interest from the science community due to the current critical energy and environmental issues. The performance of a thermoelectric material is quantified by the dimensionless figure of merit $ZT=S^2\sigma T/(\kappa_p+\kappa_e)$, where $S$ is the Seebeck coefficient, $\sigma$ is the electrical conductivity, $T$ is the absolute temperature, $\kappa_p$ and $\kappa_e$ are the electronic and lattice thermal conductivity, respectively. A good thermoelectric material should have large Seebeck coefficient and electrical conductivity, and/or low thermal conductivity. It is challenging to achieve a high $ZT$ value since optimizing one transport coefficient often leads to another adversely affected, which hinders the wide applications of thermoelectric materials. It is very important to find methods to solve such problem.

Nanostructuring\cite{MS1-1993,MS2-1993} and band convergence\cite{O.Rabin-2001} were suggested to be two promising solutions. The two- or one-dimensional structures could have much larger $ZT$ values than their bulk counterparts, due to the enhanced power factor ($PF=S^2\sigma$) caused by a sharper density of states near the Fermi level, or the reduced lattice thermal conductivity caused by the increased phonon scattering. On the other hand, it was found that the conduction or valence band extrema can be modulated and converged by tuning the doping and composition.\cite{Y.Pei-Nature,W.Liu-PRL,X.J.Tan-PRB} Such so called band convergence can significantly enhance the Seebeck coefficient without detrimental effects on the electrical conductivity. If we can choose an appropriate material and apply the two methods in it, good thermoelectric performance can be expected.

Recently, the single layer of black phosphorus (black-P), the so called phosphorene, was successfully synthesized.\cite{Y.B.Zhang-2014,H.Liu-ACS nano,W.Lu-Nano-research} Black-P is an elemental solid which is the most stable form among the phosphorus allotropes under normal condition.\cite{T.Nishii-1987} Similar to graphite, black-P crystallizes in a layered structure, namely, each phosphorus atom is covalently connected to three neighboring phosphorus atoms to form a puckered layer. It is a direct-gap semiconductor with a band gap of about 0.33 eV.\cite{R.W.Keyes-1953} When the black-P is exfoliated into few or even single layer, extraordinary optoelectronic and electronic properties emerge.\cite{Y.B.Zhang-2014,H.Liu-ACS nano,F.Xia-arXiv,J.Qiao-arXiv,D.F.Shao-arXiv,V.Tran-arXiv} A much larger direct band gap of about 2 eV was found in monolayer phosphorene.\cite{V.Tran-arXiv} In particular, the nanostructuring can largely enhance the thermoelectric performance: the $ZT$ value of monolayer phosphorene is much larger than that of bulk black-P.\cite{H.Y.Lv-arXiv,R.Fei-arXiv} This implies the nanoscale phosphorene-based material can be a good candidate of the applicable thermoelectric material. Very recently, it was reported that the electronic band structure of phosphorene can be tuned by the in-plane\cite{R.Fei-Nanolett,X.Peng-arXiv} or out-of-plane strains.\cite{A.S.Rodin-PRL} This inspires us to investigate whether the band convergence can be introduced into the system by the method of strain. If so, the thermoelectric performance should be further optimized.

In this work, we demonstrate that we can direct the conduction band convergence of phosphorene under a simple tensile strain condition, which will greatly enhance the thermoelectric performance of phosphorene. The effect of the uniaxial strain on the thermoelectric properties of phosphorene is investigated using the first-principles calculations combined with the semi-classical Boltzmann transport theory. Our results show that with a moderate tensile strain, the conduction band extrema of phosphorene can be converged, which results in an increase in the Seebeck coefficient. At the same time, the electrical conductivity at a particular direction is dramatically increased and therefore the largely increased power factor is obtained. When the zigzag-direction strain is applied, the largest $ZT$ value of 1.65 at 300 K is obtained in the zigzag direction of phosphorene at the critical strain of 5\%, conservatively estimated by using the bulk lattice thermal conductivity. The room temperature $ZT$ value can reach 2.12 in the armchair direction of phosphorene under an 8\% armchair-direction strain.

\section{Computational details}

The structural and electronic properties of phosphorene are investigated using the first-principles pseudopotential method as implemented in the ABINIT code.\cite{X.Gonze-2009,X.Gonze-2002,X.Gonze-2005} The Brillouin zones are sampled with a $10 \times1\times8$ Monkhorst-Pack $k$-mesh. The cutoff energy for the plane wave expansion is set to be 800 eV. For the structural relaxation, the exchange-correlation energy is in the form of Perdew-Burke-Ernzerhof (PBE)\cite{PBE-1996} with generalized gradient approximation (GGA). Both the geometries and atomic positions are fully relaxed until the force acting on each atom is less than $0.5 \times 10^{-3}\,{\mbox{eV}}/{\mbox{\AA}}$. In the calculations of the electronic structures, the TB-mBJ potential\cite{TB-mBJ} is used, which can reproduce accurate band gaps for many semiconductors. Based on the electronic structure, the electronic transport coefficients are derived by using the semi-classical Boltzmann theory within the relaxation time approximation\cite{Boltz} and doping is treated by the rigid band model.\cite{Rigid-band} The electronic thermal conductivity $\kappa_e$ is calculated using the Wiedemann-Franz law $\kappa_e=L\sigma T$, where $L$ is the Lorenz number. In this work, we use the Lorenz number of $1.5\times10^{-8}\,\mbox{W}\Omega/\mbox{K}^2$.\cite{Franz-law}

\section{RESULTS AND DISCUSSION}

Figure 1 shows the structure of phosphorene, with the top and side views illustrated in Figs. 1(a) and (b), respectively. The dashed line denotes the primitive cell used in our calculation, with the corresponding first Brillouin zone inserted in the figure. To confirm the reliability of our method, we first do the calculation for bulk black-P, which has the experimental results to compare with. The van der Waals interactions between the neighboring layers of bulk black-P are treated by the vdW-DFT-D2 functional.\cite{vdW-DFT} The calculated lattice parameters are $a$=3.34 {\AA}, $b$=10.51 {\AA} and $c$=4.43 {\AA}, which are very close to the experimental values.\cite{L.Cartz-1979} The bulk black-P is semiconducting with a direct band gap of 0.34 eV at the $Z$ point, in good agreement with those reported experimentally.\cite{R.W.Keyes-1953,D.Warschauer-1963} Our calculation confirms that the TB-mBJ potential can accurately predict the band gap of our investigated system, which is an important factor in determining the electronic transport properties. In the following, we use the same method to deal with the calculations for phosphorene. The lattice parameters of phosphorene are calculated to be $a_1$=3.32 {\AA} and $a_2$=4.63 {\AA}. The strain is applied along the zigzag or armchair direction of phosphorene, as indicated in Fig. 1(a).

\begin{figure}
\includegraphics[width=0.8\columnwidth]{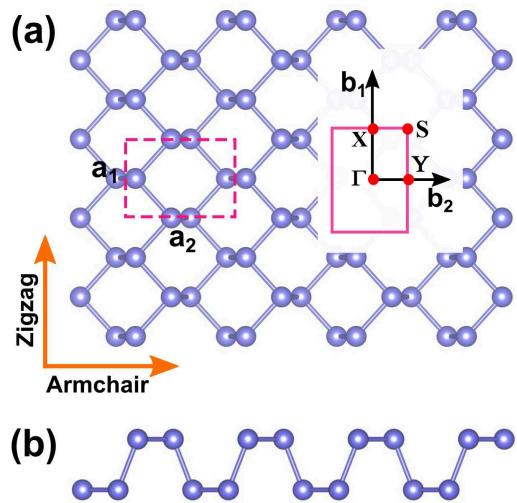}\caption{\label{fig1-structure}The top (a) and side (b) views of phosphorene. The dashed rectangle denotes the primitive cell, with the corresponding first Brillouin zone inserted.}
\end{figure}

\subsection{Strain applied along the zigzag direction}

\subsubsection{Energy band structure}

\begin{figure*}
\includegraphics[width=1.5\columnwidth]{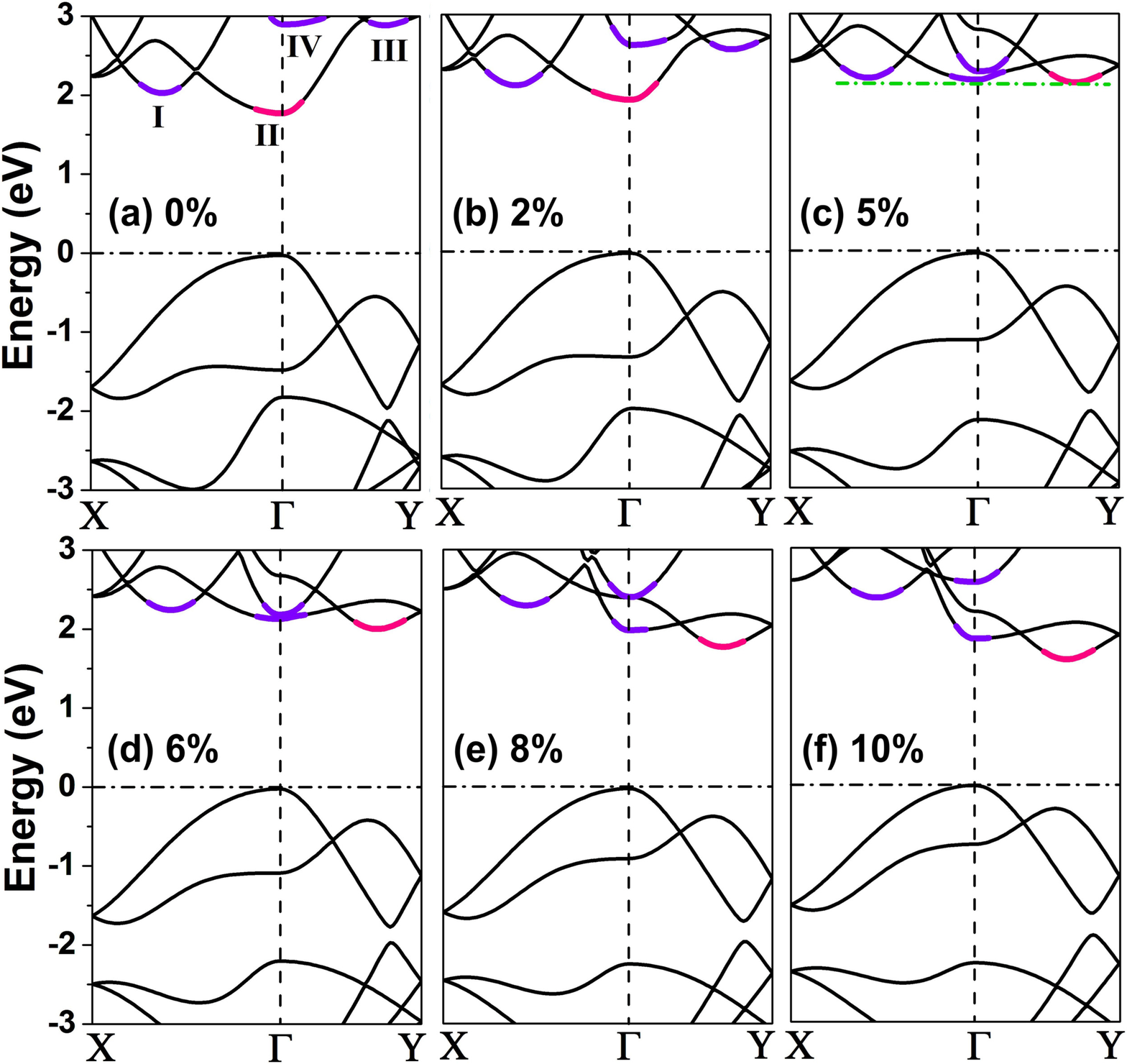}\caption{\label{fig2-energy band structure}Band structures of phosphorene under the uniaxial strain applied along the zigzag direction.}
\end{figure*}

First, we discuss the case when the tensile strain is applied along the zigzag direction.

Figure 2 shows the evolution of the band structure of phosphorene under the uniaxial strain along the zigzag direction. When no strain is applied (see Fig. 2(a)), the phosphorene is a direct-band-gap semiconductor with a gap of 1.80 eV located at the $\Gamma$ point. The band gap is much larger than that calculated using DFT-PBE method (0.92 eV)\cite{R.Fei-Nanolett} but is very close to the value calculated by the $GW$ method (2.0 eV).\cite{V.Tran-arXiv} The conduction band minimum (CBM) of phosphorene is highlighted in the red color. Note that there exist the other three band extrema located at the $\Gamma$ point and along the $\Gamma$-$X$ and $\Gamma$-$Y$ directions, respectively, highlighted in the blue color. The four band extrema are denoted by the symbols ``\Rmnum{1}", ``\Rmnum{2}", ``\Rmnum{3}", and ``\Rmnum{4}" respectively. The applied zigzag-direction strain mainly affects the conduction bands of phosphorene. As increasing the strain, the band extremum ``\Rmnum{3}" is gradually lowered while the band extremum ``\Rmnum{1}" is elevated gradually. As for the band extrema ``\Rmnum{2}" and ``\Rmnum{4}", there exists a critical strain, that is, 5\%. When the strain is smaller than 5\%, the band extremum ``\Rmnum{2}" is elevated while ``\Rmnum{4}" is lowered as increasing the value of strain. The band extrema ``\Rmnum{2}" and ``\Rmnum{4}" reach their maximum and minimum respectively at the strain of 5\%. When further increasing the strain, the band extremum ``\Rmnum{2}" is however lowered and ``\Rmnum{4}" is elevated gradually. The different behavior of the four band extrema leads to the band convergence at the strain of 5\%, which is denoted by the dashed green line in Fig. 2(c). The convergence of the band extrema will in turn lead to the increase in the Seebeck coefficient, which will be discussed later. Moreover, when the strain is smaller than 5\%, the CBM of phosphorene locates at the $\Gamma$ point (see Figs. 2(a) and (b)); when the strain reaches 5\%, the band extremum ``\Rmnum{3}" becomes energetically lower than band extrema ``\Rmnum{1}", ``\Rmnum{2}", and ``\Rmnum{4}", and thus the transition of direct-indirect band gap occurs. When the doping level in the system is low, the electrical conductivity is dominated by the CBM, so we can expect that at the critical strain of 5\%, the electrical transport property will be changed dramatically. We will come back to this point later.

\subsubsection{Electronic transport coefficients}

Based on the calculated band structure, the electronic transport coefficients of phosphorene can be evaluated by using the semi-classical Boltzmann theory and rigid band model. To get reliable results, a very dense $k$ mesh up to 840 points in the irreducible Brillouin zone (IBZ) was used. Within this method, the Seebeck coefficient $S$ can be calculated independent of the relaxation time $\tau$; however, the electrical conductivity $\sigma$ can only be calculated with $\tau$ inserted as a parameter, that is, what we obtain is $\sigma$/$\tau$. The relaxation time $\tau$ is determined by the formula $\mu=e\tau/m^*$, where $\mu$ is the carrier mobility and $m^*$ is the band effective mass. The effective mass tensor $m_{ij}^*$ is defined as $m_{ij}^*=\hbar^2[\partial^2\varepsilon(k)/\partial k_i\partial k_j]^{-1}$.

The mobility $\mu$ of phosphorene is calculated using the deformation potential (DP) theory on the basis of the effective mass approximation:\cite{J.Bardeen-1950,P.J.Price-1981,J.Xi-2012}
\begin{equation}\label{1}
    \mu=\frac{2e\hbar^3C}{3k_BT|m^*|^2E_1^2},
\end{equation}
where $T$ is the temperature. $C$ is the elastic modulus and for the 2$D$ system, the in-plane value is defined as $C^{2D}=[\partial^2 E/\partial\delta^2]/S_0$, where $E$, $\delta$ and $S_0$ are, respectively, the total energy, the applied uniaxial strain and the area of the investigated system. The DP constant $E_1$ along a certain direction is obtained by $E_1=dE_{edge}/d\delta$, where $E_{edge}$ is the energy of the band edges (valence band maximum for the holes and conduction band minimum for the electrons) and $\delta$ is the applied small strain (by a step of 0.5\%).

\begin{figure}
\includegraphics[width=1.0\columnwidth]{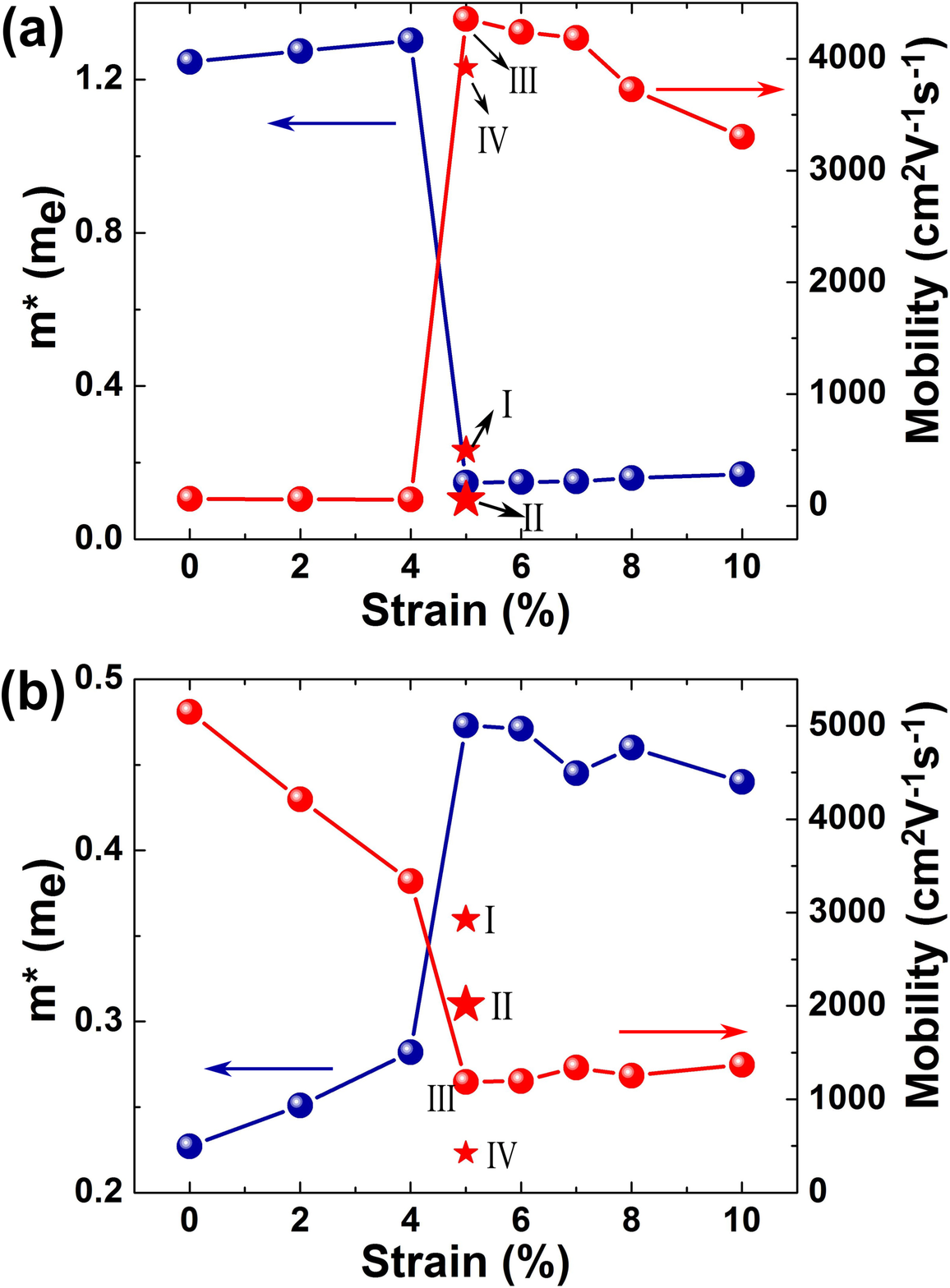}\caption{\label{fig3-mobility} Effective mass $m^*$ and carrier mobility $\mu$ of phosphorene at 300 K along (a) the zigzag and (b) the armchair directions as a function of the applied strain along the zigzag direction. At the critical strain of 5\%, the carrier mobility of band extrema ``\Rmnum{1}", ``\Rmnum{2}" and ``\Rmnum{4}" are also denoted by the red stars.}
\end{figure}

The elastic modulus $C$ can be obtained by fitting the curve of energy-strain relationship. The calculated value of $C$ in the zigzag direction is 106.18 N/m. The value is smaller than that of $\mbox{MoS}_2$ (about 120 N/m)\cite{R.C.Cooper-2013} and graphene (about 335.0 N/m).\cite{C.Lee-2008} The DP constant $E_1$ is obtained by the linear fitting of the energy shift of band edges as a function of the strain. The calculated $E_1$ is 3.98 eV for electrons. When the uniaxial zigzag-direction strain is applied, the effective mass $m^*$ as well as the corresponding mobility $\mu$ at 300 K in the zigzag and armchair directions are shown in Figs. 3(a) and (b), respectively. Since the strain mainly modulates the bands above the Fermi level, we only focus on the properties of electrons. We can see that for the zigzag direction (see Fig. 3(a)), there is a sharp drop of the effective mass $m^*$ at the critical strain (5\%), which in turn leads to the dramatically increased carrier mobility, with the values about two orders of magnitude larger than those before the critical strain. Note that here the effective mass and mobility are calculated with respect to the CBM (highlighted in red color in Fig. 2). However, at the critical strain of 5\%, the energy difference of the four conduction band extrema is very small, thus except for the CBM, the other three band extrema will also contribute to the electrical conductivity. The values of the carrier mobility of band extrema ``\Rmnum{1}", ``\Rmnum{2}" and ``\Rmnum{4}" are denoted by the red stars in the Fig. 3. The larger the size of the star, the more it will contribute to the electrical conductivity. After the critical strain, the greatly enhanced mobility and therefore the electrical conductivity will be very beneficial to the thermoelectric application. The trend of the mobility along the armchair direction (see Fig. 3(b)) is just reversed. However, although there is a sharp drop of the mobility at the critical strain, the mobilities after that point are still considerable, larger than 1000 $\mbox{cm}^2\mbox{V}^{-1}\mbox{s}^{-1}$. The different behavior of the effective mass and carrier mobility in the two different directions upon the applied zigzag-direction strain will in turn influence their thermoelectric properties, which will be discussed later.

\begin{table*}

\caption{\label{Table1-relaxation time}Effective mass ($m^*$), carrier mobility ($\mu$) and relaxation time ($\tau$) at 300 K in the zigzag and armchair directions of phosphorene under the uniaxial zigzag-direction strain. At the critical strain of 5\%, only the values of the CBM are listed.}

\begin{tabular}{ccccccccccccccccccc}
\hline
\hline
~&~& Strain &~& 0\% &~& 2\% &~& 4\% &~& 5\% &~& 6\% &~& 7\% &~& 8\% &~& 10\%\tabularnewline
\hline
Zigzag &~& $m^*$($m_e$) &~& 1.246 &~& 1.274 &~& 1.302 &~& 0.148 &~& 0.150 &~& 0.151 &~& 0.160 &~& 0.170\tabularnewline
~&~& $\mu$($\mbox{cm}^2\mbox{V}^{-1}\mbox{s}^{-1}$) &~& 61.46 &~& 58.79 &~& 56.24 &~& 4356.09 &~& 4240.65 &~& 4184.72&~& 3727.14&~& 3301.55\tabularnewline
~&~& $\tau$(fs) &~& 43.53 &~& 42.57 &~& 41.62 &~& 366.45 &~& 361.56 &~& 359.17 &~& 338.96 &~& 319.02\tabularnewline
\hline
Armchair &~& $m^*$($m_e$) &~& 0.227 &~& 0.251 &~& 0.282 &~& 0.473 &~& 0.471 &~& 0.445 &~& 0.460 &~& 0.440\tabularnewline
~&~& $\mu$($\mbox{cm}^2\mbox{V}^{-1}\mbox{s}^{-1}$) &~& 5149.24 &~& 4211.61 &~& 3336.55 &~& 1185.97 &~& 1196.06 &~& 1339.91 &~& 1253.95 &~& 1370.53\tabularnewline
~&~& $\tau$(fs) &~& 664.39 &~& 600.86 &~& 534.81 &~& 318.85 &~& 320.20 &~& 338.91 &~& 327.86 &~& 342.76\tabularnewline
\hline
\hline
\end{tabular}

\end{table*}

Based on the calculated effective mass $m^*$ and carrier mobility $\mu$, we are now able to obtain the relaxation time $\tau$. In Table I, we summarize the relaxation time $\tau$ at 300 K along the two different (zigzag and armchair) directions  of phosphorene when the uniaxial zigzag-direction strain is applied, with the corresponding effective mass $m^*$ and carrier mobility $\mu$ included. We can see that before the critical strain of 5\%, the relaxation times along the armchair direction are much larger than those along the zigzag direction. However, they become comparable when the strain reaches 5\%.

\begin{figure*}
\includegraphics[width=1.6\columnwidth]{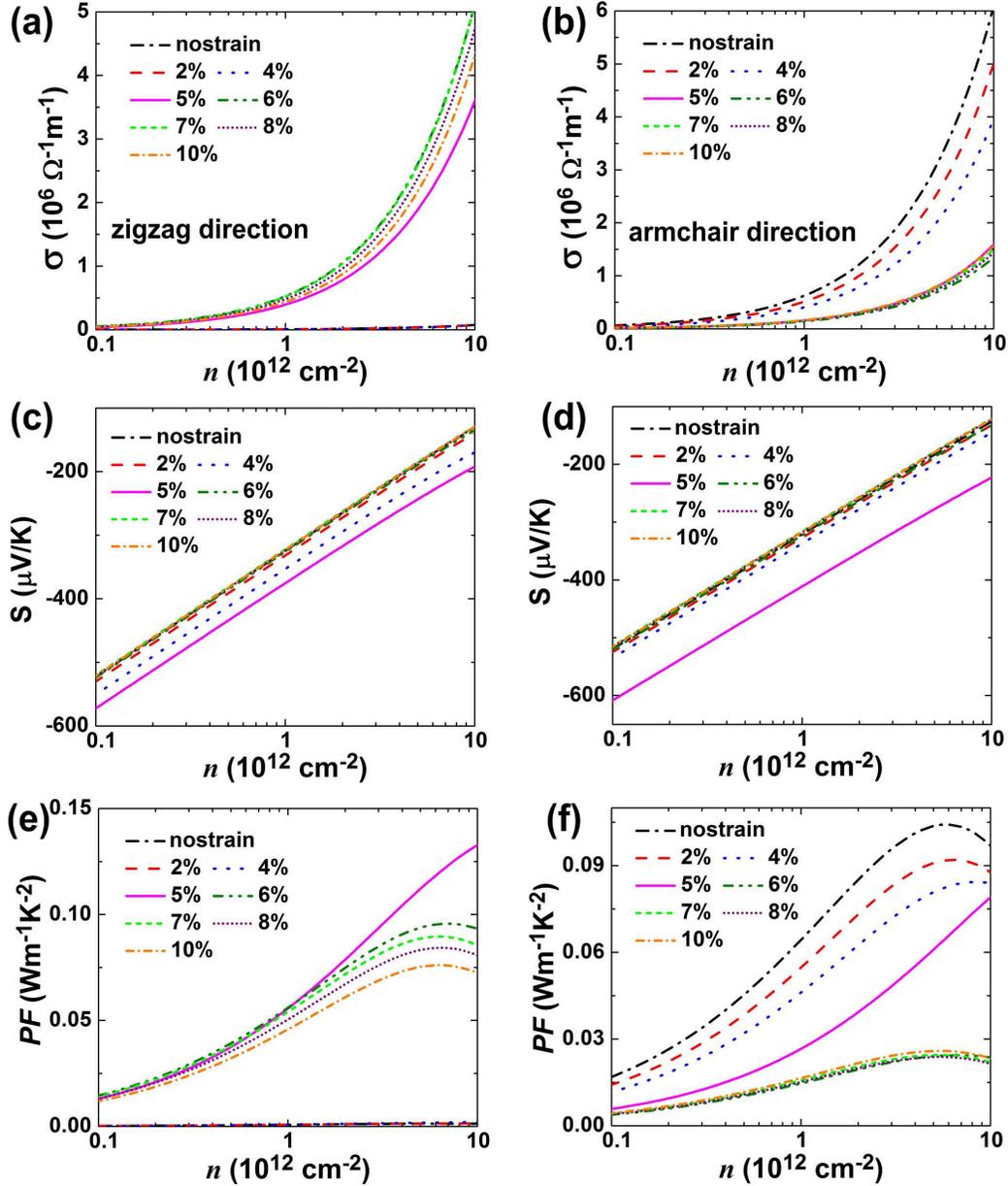}\caption{\label{fig4-electronic transport coefficients} Carrier concentration dependence of the electronic transport coefficients at 300 K when the uniaxial strain is applied along the zigzag direction: (a)electrical conductivity along the zigzag direction; (b)electrical conductivity along the armchair direction; (c) Seebeck coefficient along the zigzag direction; (d)Seebeck coefficient along the armchair direction; (e) power factor ($PF$) along the zigzag direction; (f) power factor ($PF$) along the armchair direction.}
\end{figure*}

Inserting the calculated relaxation time $\tau$, we plot in Figs. 4(a) and (b) the room-temperature electrical conductivity $\sigma$ as a function of the carrier concentration $n$ along the zigzag and armchair directions, respectively, when the different zigzag-direction strains are applied. When no strain is applied, the electrical conductivity exhibits obvious anisotropic property, with the value along the armchair direction much larger than that along the zigzag direction. This anisotropic property is due to the different dispersions of the CBM along the $\Gamma$-$X$ (zigzag direction in the real space) and $\Gamma$-$Y$ (armchair direction in the real space) directions. The band along the $\Gamma$-$X$ direction is much flatter than that along the $\Gamma$-$Y$ direction, which results in the much larger band effective mass and therefore much smaller carrier mobility and electrical conductivity in the zigzag direction. For the transport along the zigzag direction, after the critical strain (5\%), the electrical conductivity $\sigma$ is obviously enhanced. If we notice the band structure of phosphorene under the critical strain (Fig. 2(c)), we can see that the conduction band minimum (the red line) which dominates the electrical conductivity moves from the $\Gamma$ point to the one along the $\Gamma$-$Y$ direction. The effective mass of that band in the zigzag direction is dramatically decreased (see Fig. 3(a) and Table I), which results in the largely increased electrical conductivity. However, in the strain range of $5\%-10\%$, the electrical conductivity under the strain of 5\% is the smallest (see Fig. 4(a)), which is consistent with our explanation above that although the mobility of the CBM is the largest at the 5\% strain, the other three band extrema ``\Rmnum{1}", ``\Rmnum{2}" and ``\Rmnum{4}" will also contribute to the electrical conductivity, which will more or less decrease $\sigma$ at that strain. The electrical conductivity along the armchair direction is however decreased by the strain due to the increased effective mass and therefore the decreased carrier mobility in this direction, as shown in Fig. 3(b).

Figures 4(c) and (d) show the Seebeck coefficient $S$ at 300 K as a function of the carrier concentration along the zigzag and armchair directions, respectively, under the zigzag-direction strain condition. For both directions, the absolute values of the Seebeck coefficients reach the maxima at the critical strain of 5\%, where the band convergence is achieved as indicated in Fig. 2(c). A high Seebeck coefficient is often caused by a high overall density-of-states effective mass $m_d^*$. The $m_d^*$ is related to the band effective mass $m^*$ (i.e., the effective mass of a single carrier pocket) by $m_d^*=N_v^{2/3}m^*$, where $N_v$ is the number of the degenerated band valleys. A large value of $N_v$ will lead to a large $m_d^*$, and therefore a large absolute value of the Seebeck coefficient. When the band extrema of the multiple bands have no or little difference in energy (i.e., the bands are converged), they are considered to be effectively degenerated and $N_v$ is the increased. Here, we obtain the band convergence by applying a uniaxial strain and a significantly increased Seebeck coefficient is achieved at the critical strain. It has been reported that the convergence of many valleys in bulk materials could be achieved by tuning the doping and composition.\cite{Y.Pei-Nature, W.Liu-PRL,X.J.Tan-PRB} Note here that for the zigzag direction and at the critical strain, not only the Seebeck coefficient reach the maximum value, but also the electrical conductivity is greatly increased, which will benefit the power factor of phosphorene.

The $PF$ as a function of the carrier concentration in the zigzag and armchair directions of phosphorene at 300 K are plotted in Figs. 4(e) and (f), respectively, when the zigzag-direction strain is applied. We can see that for the zigzag direction (Fig. 4(e)), the $PF$ of phosphorene is greatly enhanced when the strain reaches 5\%. At relatively higher carrier concentration, the phosphorene under 5\% zigzag-direction strain has the largest $PF$, due to the largest Seebeck coefficient $S$ and greatly enhanced electrical conductivity $\sigma$. However, for the armchair direction, although the phosphorene under 5\% strain has the largest Seebeck coefficient, the electrical conductivity is greatly decreased, which leads to the decrease of the $PF$. The electrical conductivity exhibits the obvious anisotropic property, which results in the different behavior of the $PF$ between the zigzag and armchair directions upon the applied zigzag-direction strain.

\subsubsection{Dimensionless figure of merit $ZT$}

\begin{figure}
\includegraphics[width=0.9\columnwidth]{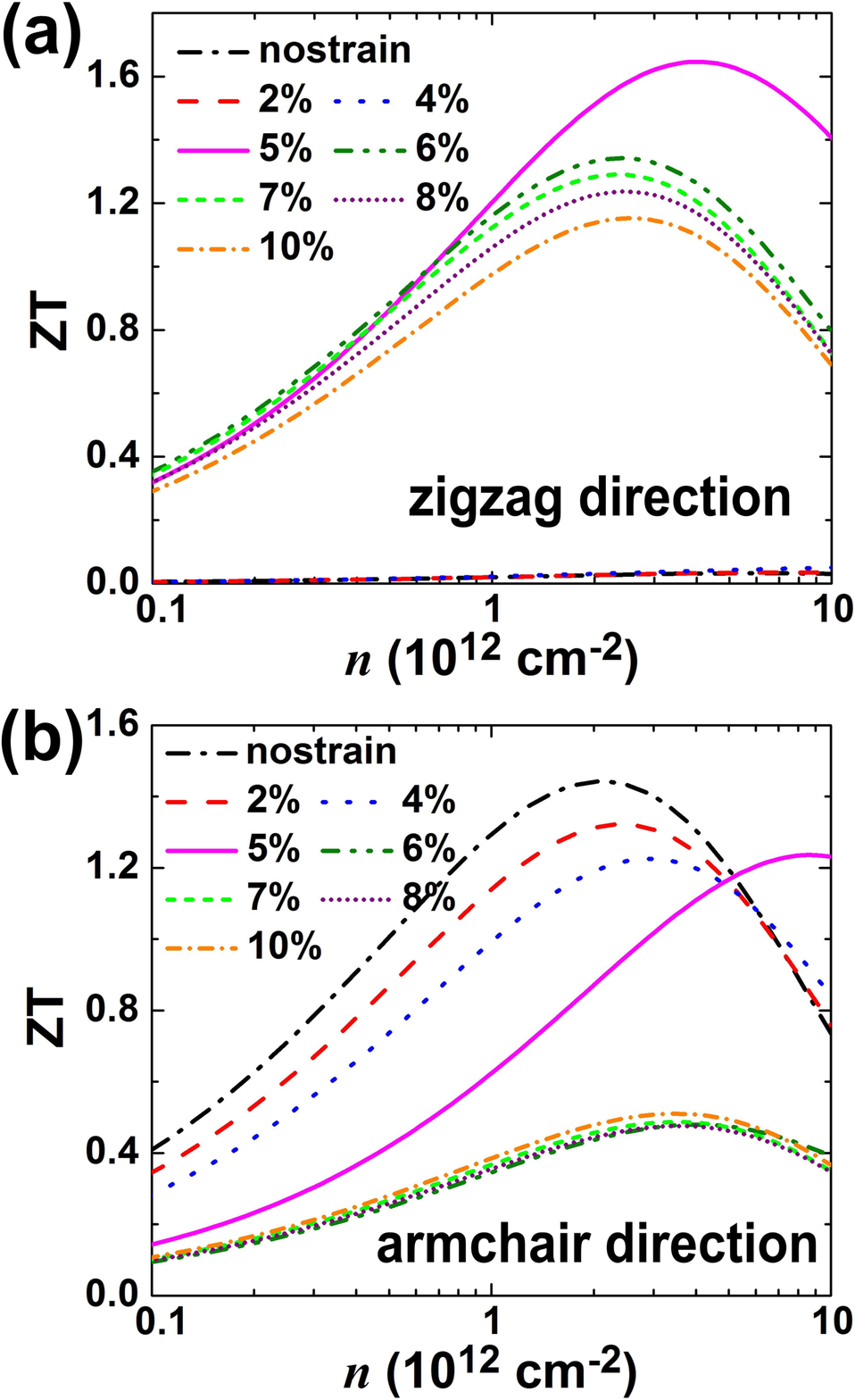}\caption{\label{fig5-ZT value} Dimensionless figure of merit $ZT$ as a function of carrier concentration at 300 K in (a) zigzag and (b) armchair direction under different uniaxial strain applied along the zigzag direction.}
\end{figure}

The electronic thermal conductivity $\kappa_e$ of phosphorene is calculated based on the Wiedemann-Franz law $\kappa_e=L\sigma T$ as mentioned above. We then estimate the $ZT$ values of phosphorene at 300 K under the zigzag-direction strain by using the experimental lattice thermal conductivity of bulk black-P (12.1 $\mbox{Wm}^{-1}\mbox{K}^{-1}$). The $ZT$ values at 300 K along the zigzag and armchair directions are plotted as a function of the carrier concentration in Figs. 5(a) and (b), respectively. We can see that the $ZT$ value of phosphorene without strain exhibits the strong anisotropic property, with the value along the armchair direction much larger than that along the zigzag direction. The applied zigzag-direction strain has different effect on the $ZT$ values in the two particular directions. For the armchair direction (see Fig. 5(b)), the $ZT$ value is decreased by the strain. The largest $ZT$ value of 1.44 can be obtained in this direction when no strain is applied and the corresponding carrier concentration is $2.06\times10^{12}\,\mbox{cm}^{-2}$. For the transport along the zigzag direction, however, the $ZT$ value is greatly enhanced by the strain. The largest $ZT$ value of 1.65 is achieved under the zigzag-direction strain of 5\%, with the corresponding carrier concentration of $3.99\times10^{12}\,\mbox{cm}^{-2}$. The maximum of the $ZT$ value is 50 times larger than that without strain. Applying strain is shown to be an effective way to tune the band structure of phosphorene and the thermoelectric performance in particular direction can be largely optimized. Note here that we use the lattice thermal conductivity of bulk black-P to estimate the $ZT$ values of phosphorene. If the lattice thermal conductivity of the phosphorene can be reduced compared with the bulk value, which can be realized in many low-dimensional structures, the $ZT$ value of the phosphorene can be further enhanced.

\subsection{Strain applied along the armchair direction}

\begin{figure}
\includegraphics[width=0.9\columnwidth]{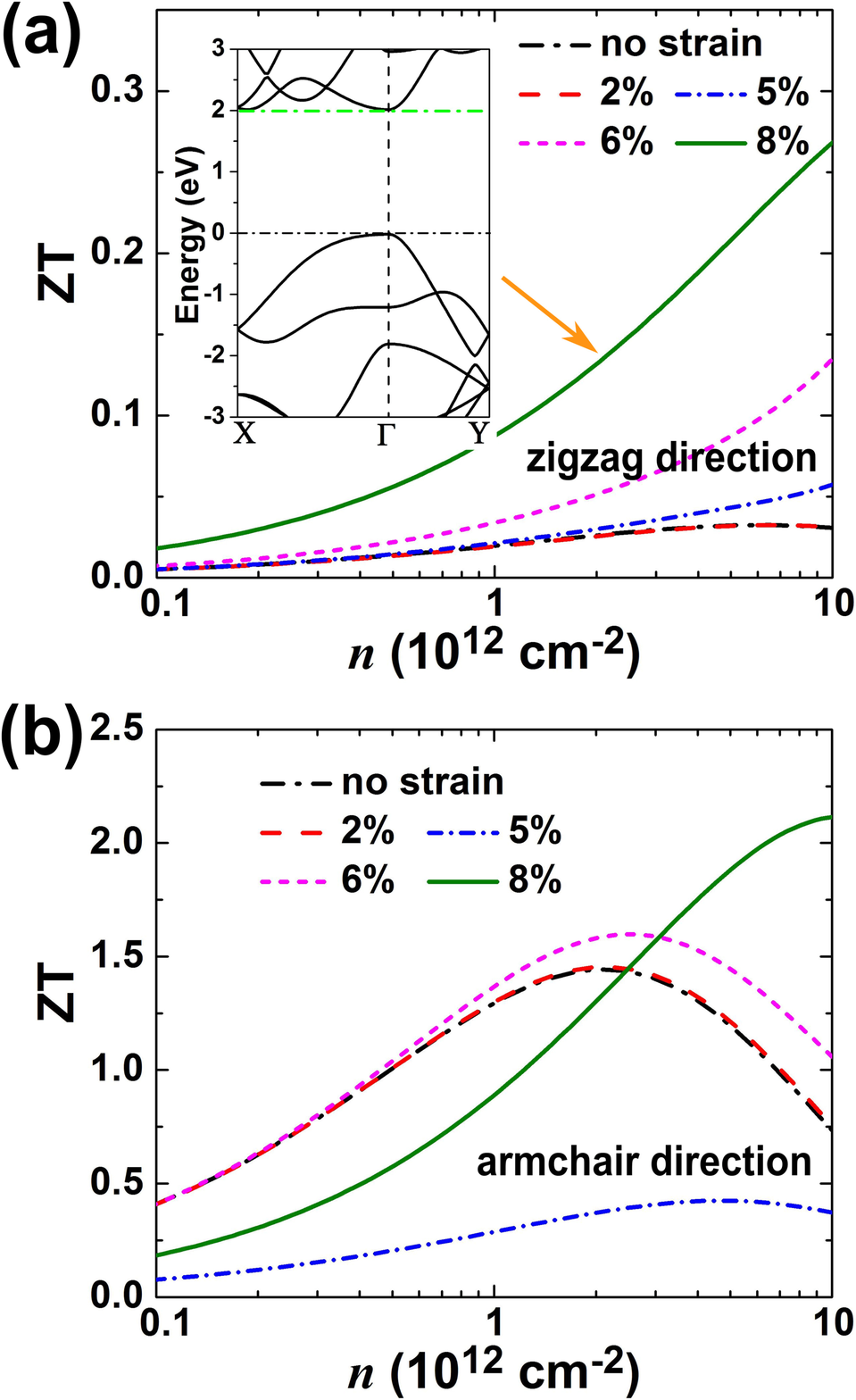}\caption{\label{fig6-ZT value} Dimensionless figure of merit $ZT$ as a function of carrier concentration at 300 K in (a) zigzag and (b) armchair direction under uniaxial strain applied along the armchair direction.}
\end{figure}

We also apply the strain along the armchair direction of phosphorene. The same as the case when the zigzag-direction strain is applied, the band structure and the thermoelectric properties of phosphorene can also be tuned by the strain. In this case, when the armchair-direction strain reaches 8\%, the band extrema convergence can be achieved, as indicated by the dashed green line in the inset in Fig. 6(a). However, the number of the converged bands is less than that when the zigzag-direction strain is applied. As a result, the enhancement of the $ZT$ value is not as large as the case when the stain is applied along the zigzag direction. The largest $ZT$ values at 300 K in the zigzag and armchair directions are respectively 7.4 and 0.5 times larger than the corresponding values when no strain is applied. At the critical strain of 8\%, we can obtain the largest room-temperature $ZT$ values of 0.27 and 2.12 in the zigzag and armchair directions of phosphorene, respectively.

\section{Conclusion}

In summary, we have investigated the strain effect on the electronic and thermoelectric properties of phosphorene based on the the first-principles calculations combined with the semi-classical Boltzmann theory. The band structure of phosphorene can be modulated by the uniaxial strain and conduction band extrema are effectively converged at the critical strain. The Seebeck coefficient is largely enhanced due to the band convergence. The electrical conductivity exhibits anisotropic property and behaves differently in the zigzag and armchair directions of phosphorene upon the applied strain. When the zigzag-direction strain is applied, the Seebeck coefficient and electrical conductivity in zigzag direction can be greatly enhanced simultaneously at the critical strain of 5\%. The largest $ZT$ value of 1.65 is then achieved, which is 50 times larger than that without strain. When the armchair-direction strain of 8\% is applied, we can obtain the largest $ZT$ value of 2.12 in the armchair direction of phosphorene. Our results indicate that strain-induced band convergence could be an effective method to enhance the thermoelectric performance of phosphorene.

\section{Acknowledgement}

This work was supported by the National Key Basic Research under Contract No.2011CBA00111, the National Nature Science Foundation of China under Contract No.11274311, the Joint Funds of the National Natural Science Foundation of China and the Chinese Academy of Sciences' Large-scale Scientific Facility (Grand No.U1232139), and Anhui Provincial Natural Science Foundation under Contract No.1408085MA11. The calculation was partially performed at the Center for Computational Science, CASHIPS.

\end{document}